\documentstyle[eqsecnum,manuscript,aps,psfig]{revtex}
\begin{document}
\title{G\"odel Universes in String Theory}
\author{John D. Barrow$^1$ and Mariusz P. D\c abrowski$^{1,2}$}
\address{$^1${\it Astronomy Centre, University of Sussex, Falmer, Brighton }\\
BN1 9QJ, U.K.,\\
$^2$Institute of Physics, University of Szczecin, 70-451 Szczecin, Poland.}
\date{\today}
\maketitle

\begin{abstract}
We show that homogeneous G\"odel spacetimes need not contain closed timelike
curves in low-energy-effective string theories. We find exact solutions for
the G\"odel metric in string theory for the full $O(\alpha ^{\prime })$
action including both dilaton and axion fields. The results are valid for
bosonic, heterotic and super-strings. To first order in the inverse string
tension $\alpha ^{\prime }$, these solutions display a simple relation
between the angular velocity of the G\"odel universe, $\Omega ,$ and the
inverse string tension of the form $\alpha ^{\prime }=1/\Omega ^2$ in the
absence of the axion field. The generalization of this relationship is also
found when the axion field is present.
\end{abstract}


PACS number(s): 98.80.Hw, 04.50.+h, 11.25.Mj, 98.80.Cq

\newpage

\section{Introduction}

\setcounter{equation}{0}

The relationship between micro and macrophysics is most commonly sought
within a framework of unified descriptions of the fundamental forces of
Nature that is provided by superstring theories or some $M$ theory believed
to underlie them \cite{witten},\cite{fradcall,zeroalfa,firstalfa}. The most
frequently investigated theory is the bosonic effective action leading to
the pre-big-bang superinflationary scenario \cite{strmod}, although some
type IIB superstring models \cite{IIB} and $M$-theory solutions have also
been found \cite{M}. However, the cosmological solutions of low-energy
string theory have been confined to solutions of the zero-order equations, 
\cite{zero}, \cite{homog}. In this Letter, we give some new solutions of
string theories, including terms up to first order in the inverse string
tension $\alpha ^{\prime }$ and both dilaton and axion fields, for
homogeneous spacetimes of the form first investigated in general relativity
by G\"odel in 1949 \cite{godel}. We recall that G\"odel's solutions
attracted considerable interest because they described rotating universes
that possessed the completely unexpected property of closed timelike curves
(CTCs)\footnote{%
It is interesting to note that Weyl, in his book {\it Space, Time and Matter,%
}\cite{weyl} foresaw that Einstein's theory might permit time travel to
occur, remarking that {\it '[in general relativity] it is not impossible for
a world-line (in particular, that of my body), although it has a timelike
direction at every point, to return to the neighbourhood of a point which it
has already passed through....[but] the very considerable fluctuations of
the [spacetime metric] would be necessary to produce this effect do not
occur in the region of the world in which we live}'\cite{weyl},\cite{imposs}.%
}. This discovery led to a reappraisal of thinking about Mach's 'Principle' 
\cite{godel1}, inspired the rigorous investigation of the possibility of
time travel in relativistic theories of gravity \cite{nahin}, and stimulated
the serious investigation of the global structure of spacetimes \cite{metric}%
. However, we find that in string theory, unlike in general relativity,
G\"odel universes need not contain CTCs. Previously, causal G\"odel
universes have been found in general relativity in the presence of massless
scalar fields by Rebou\c cas and Tiomno\cite{tiomno}, and in gravity
theories derived from an action containing terms quadratic in the Ricci
curvature invariants by Accioly \cite{accio}. We find that CTCs need not
occur in the G\"odel universes in heterotic and bosonic string theories
containing the full bosonic particle spectrum of axion, dilaton, and
graviton.

\section{Low-energy-effective $O(\alpha ^{\prime })$ action for strings}

\setcounter{equation}{0}

We write the general form of the string effective action to $\alpha ^{\prime
}$ order in the string frame as, \cite{firstalfa}, 
\begin{eqnarray}
S &=&\int d^nx\sqrt{-g}~e^{-\phi }\left\{ R-2\Lambda +(\partial \phi )^2-%
\frac 1{12}H^2\right.   \nonumber \\
&&\ -\alpha ^{\prime }\lambda _0\left[ R_{\mu \nu \sigma \rho }R^{\mu \nu
\sigma \rho }-\frac 12R^{\mu \nu \sigma \rho }H_{\mu \nu \alpha }H_{\sigma
\rho }{}^\alpha +\right.   \nonumber \\
&&\ \left. \left. \frac 1{24}H_{\mu \nu \lambda }H^\nu {}_{\rho \alpha
}H^{\rho \sigma \lambda }H_\sigma {}^{\mu \alpha }-\frac 18H_{\mu \rho
\lambda }H_\nu {}^{\rho \lambda }H^{\mu \sigma \alpha }H^\nu {}_{\sigma
\alpha }\right] +O(\alpha ^{\prime }{}^2)\right\}   \label{ts}
\end{eqnarray}
where $\lambda _0=-\frac 18$ for heterotic strings, $-\frac 14$ for bosonic
strings and $0$ for superstrings; $n$ is the number of spacetime dimensions, 
$\alpha ^{\prime }$ is the inverse string tension parameter, and the action
is truncated at first order. Thus, in the case of superstrings the leading
corrections after the zero-order terms will enter at fourth order and will
not be investigated in this paper. These higher-order terms lead to a
significant increase in algebraic complexity. In (2.1) $\phi $ is the
dilaton, $g$ the determinant of the metric, $\Lambda $ the cosmological
constant, $R$ the Ricci scalar, $R_{\mu \nu \rho \sigma }$ the Riemann
tensor, and $H_{\mu \nu \rho }=\partial _{[\mu }B_{\nu \rho ]}$ is the axion
with $H^2=H_{\mu \nu \rho }H^{\mu \nu \rho }$ where $B_{\mu \nu }$ is the
antisymmetric tensor potential. The field equations for bosonic strings then
take the following form 
\begin{eqnarray}
R_{\mu \beta } &+&\partial _\mu \partial _\beta \phi -\frac 14H_{\mu \nu
\lambda }H_\beta {}^{\nu \lambda }+\frac 12\alpha ^{\prime }\left[ R_{\mu
\nu \sigma \rho }R_\beta {}^{\nu \sigma \rho }-R_\mu {}^{\nu \sigma \rho
}H_{\beta \nu \alpha }H_{\sigma \rho }{}^\alpha +\right.   \nonumber \\
\  &&\left. \frac 18H_{\mu \nu \lambda }H^\nu {}_{\rho \alpha }H^{\rho
\sigma \lambda }H_{\sigma \beta }{}^\alpha -\frac 38H_{\mu \rho \lambda
}H_\nu {}^{\rho \lambda }H_\beta {}^{\sigma \alpha }H^\nu {}_{\sigma \alpha
}\right] =0,
\end{eqnarray}
\begin{eqnarray}
R- &&2\Lambda +2\partial _\mu \partial ^\mu \phi -\partial _\mu \phi
\partial ^\mu \phi -\frac 1{12}H^2+\frac 14\alpha ^{\prime }\left[ R_{\mu
\nu \sigma \rho }R^{\mu \nu \sigma \rho }-\frac 12R^{\mu \nu \sigma \rho
}H_{\mu \nu \alpha }H_{\sigma \rho }{}^\alpha +\right.   \nonumber \\
\  &&\left. \frac 1{24}H_{\mu \nu \lambda }H^\nu {}_{\rho \alpha }H^{\rho
\sigma \lambda }H_\sigma {}^{\mu \alpha }-\frac 18H_{\mu \rho \lambda }H_\nu
{}^{\rho \lambda }H^{\mu \sigma \alpha }H^\nu {}_{\sigma \alpha }\right] =0.
\end{eqnarray}
These equations are completed by the axion equation of motion 
\begin{equation}
\partial _\mu (\delta S/\delta (\partial _\mu B_{\nu \lambda }))=\left[
e^{-\phi }\left( H^{\mu \nu \lambda }+\alpha ^{\prime }M^{\mu \nu \lambda
}\right) \right] _{;\mu }=0,
\end{equation}
where 
\begin{equation}
M^{\mu \nu \lambda }=\frac 32R^{\mu \nu \sigma \rho }H_{\sigma \rho
}{}^\lambda +\frac 14H_\sigma {}^{\mu \alpha }H^\nu {}_{\rho \alpha }H^{\rho
\sigma \lambda }-\frac 34H^{\mu \sigma \alpha }H_\rho {}^{\nu \lambda
}H^\rho {}_{\sigma \alpha }.
\end{equation}
To zeroth order in $\alpha ^{\prime },$ eqn (2.4) is just $(e^{-\phi }H^{\mu
\nu \lambda })_{;\nu }=0$ (see e.g. \cite{homog}) where semicolon denotes a
covariant derivative with respect to the metric. The cosmological constant
term $\Lambda $ is related to the dimension of space and the inverse string
tension by \cite{zeroalfa,strmod} 
\begin{equation}
\Lambda =\frac{n-26}{3\alpha ^{\prime }}.
\end{equation}
The relation (2.6) holds for bosonic strings; for superstrings $\Lambda $ is
proportional to $n-10$.

\section{Exact Stringy G\"odel Universes}

\setcounter{equation}{0}

The G\"odel metric describes a homogeneous spacetime \cite{godel}. Its line
element in cylindrical coordinates $(t,r,z,\psi )$ is usually given by
either of the two forms \cite{tiomno} 
\begin{equation}
ds^2=-\left[ dt+C(r)d\psi \right] ^2+D^2(r)d\psi ^2+dr^2+dz^2,
\end{equation}
or 
\begin{eqnarray}
ds^2=-dt^2-2C(r)dtd\psi +G(r)d\psi ^2+dr^2+dz^2,  \nonumber
\end{eqnarray}
where the radial functions have the form 
\begin{eqnarray}
C(r) &=&\frac{4\Omega }{m^2}\sinh ^2{\left( \frac{mr}2\right) }, \\
D(r) &=&\frac 1{m^2}\sinh {(mr)}, \\
G(r) &=&\frac 4{m^2}\sinh ^2{\left( \frac{mr}2\right) }\left[ 1+\left( 1-%
\frac{4\Omega ^2}{m^2}\right) \sinh ^2{\left( \frac{mr}2\right) }\right] ,
\end{eqnarray}
with $m$ and $\Omega $ constants. In order to{\it \ avoid }the existence of
CTC's in these spacetimes we require 
\begin{equation}
G(r)=D^2(r)-C^2(r)>0.
\end{equation}
In a G\"odel universe, the four-velocity of matter is $u^\alpha =\delta
_0^\alpha $ and the rotation vector is $V^\alpha =\Omega \delta _0^3$ while
the vorticity scalar is given by $\omega =\Omega /\sqrt{2}$. The original
G\"odel metric of general relativity,\cite{godel}, has $m^2=2\Omega ^2$ and
obviously contradicts (3.5). There has been extensive discussion of the
generality and significance of the presence of CTC's in the G\"odel metric
in general relativity \cite{nahin},\cite{metric}.

The only nonvanishing components of the Riemann tensor in an orthonormal
frame permitted by the spacetime homogeneity of the G\"odel universe are
constant, \cite{accio}, with 
\begin{equation}
R_{0101}=R_{0202}=\frac 14\left( \frac{C^{\prime }}D\right) ^2=\Omega ^2,%
\hspace{0.5cm}R_{1212}=\frac 34\left( \frac{C^{\prime }}D\right) -\frac{%
D^{\prime \prime }}D=3\Omega ^2-m^2,
\end{equation}
and the prime means the derivative with respect to $r$. It is interesting to
note that with (3.6) the Gauss-Bonnet term $R_{GB}^2\equiv R_{\mu \nu \sigma
\rho }R^{\mu \nu \sigma \rho }-4R_{\mu \nu }R^{\mu \nu }+R^2$ vanishes. This
term appears in redefined $O(\alpha ^{\prime })$ actions (2.1) (see e.g. 
\cite{meiss}). Because of the metric symmetry we assume that the dilaton
depends only on the coordinate along the axis of rotation, $z$, so 
\begin{equation}
\phi =\phi (z)=fz+\phi _0,
\end{equation}
where $f$ and $\phi _0$ are constants. For the axion, a short analysis
similar to that given in \cite{homog} shows that the only possible ansatz
which is consistent with the form (3.7) for the dilaton is 
\begin{equation}
H_{012}=-H^{012}=E,
\end{equation}
with $E$ constant. This can also be expressed in terms of the pseudoscalar
axion field, $h$, \cite{ed,homog}, by 
\begin{equation}
h(z)=\frac Ef\exp {(-fz-\phi _0)}+h_0.
\end{equation}
The ans\"atze of eqs. (3.7)-(3.8) guarantee that the axion's equation of
motion is satisfied to zeroth order in $\alpha ^{\prime }$. Since the axion
field's 3-form strength is defined by an external derivative of the
antisymmetric tensor potential, $B_{\mu \nu }$, one can easily show that the
only nonvanishing component of the potential is time-dependent 
\begin{equation}
B_{12}(t)=Et+t_0.
\end{equation}
This is expected from the discussion of \cite{homog,ed} since this
restriction occurs in every case where the spacetime possesses a
distinguished direction. It should then possess $O(n-1,n-1)$ symmetry
(despite the fact the metric is not time-dependent): which is an example of $%
T$-duality.

Note the generality of the action (2.1): it is given for any spacetime
dimension $n,$ and it contains the full spectrum of graviton, axion, and
dilaton. It also possesses a general $O(n-1,n-1)$ symmetry \cite{meiss}.
Since this manifests itself in complicated forms in individual solutions, we
first discuss two special cases before giving the general G\"odel solution
for this string theory.

\subsection{Zeroth order in $\alpha ^{\prime }$}

The field equations (2.2)-(2.3) (which are in the string frame) to zeroth
order in $\alpha ^{\prime }$, together with eqs. (3.6)-(3.8) with $n=4$,
possess a G\"odel solution with metric (3.1)-(3.4) if the following
relations hold between the constants $\Omega ,m,E,f$ and $\Lambda :$ 
\begin{equation}
\Omega ^2=\frac{m^2}4=\frac{E^2}4=\frac{-2\Lambda -f^2}4.
\end{equation}
When $\Lambda =0$ the relation (3.11) has no meaning for $m^2>0$, but it can
be applied for $m^2\equiv -\mu ^2<0$, in which case there exists an infinite
sequence of both causal and acausal regions (see \cite{tiomno}). We do not
consider such solutions here.

>From the relations (3.11) one can easily deduce that the axion field must
not vanish if a solution is to exist and act as a source of rotation. This
is consistent with the known repulsive behaviour of the axion acting as a
torsion field, which leads to bouncing solutions within the string theory
(e.g. Ref.\cite{behrndt}). However, from (3.11), there is another
constraint, 
\begin{equation}
\Lambda <-\frac{f^2}2,
\end{equation}
which shows that the cosmological term has to be negative. Formally, the
dilaton can vanish without disrupting the causal structure of the solution
(3.1), but in that case the axion plays the role of a scalar field minimally
coupled to gravity - the case already studied in ref. \cite{tiomno}.

These results are quite different from the situation in general relativity
with both a scalar field and electromagnetic field present \cite{tiomno}. It
is known \cite{ed} that the action (2.1) can be transformed to the Einstein
frame where the field equations resemble Einstein gravity with source terms
given by axion and dilaton fields\cite{ed}. Using the ans\"atze (3.7)-(3.9)
we checked the validity of the formulae (6.8)-(6.9) of Ref.\cite{tiomno}, to
obtain 
\begin{equation}
\Omega ^2=\frac{m^2}4=\frac{f^2}4=-2\Lambda .
\end{equation}
We see also that (in analogy to the electromagnetic field that is present
there), we cannot admit the axion and obtain a causal model (3.1) in the
Einstein frame because there is no way to fulfil the dilaton equation of
motion. This, perhaps, shows the superiority of the string frame over the
Einstein frame, especially since the axion is a typical 'stringy' particle
which should not be excluded from the spectrum of particles. Note that $%
\Lambda $ has to be negative in both frames. For $\lambda _0=0$ in (2.1),
these conclusions also hold for superstrings.

\subsection{First order in $\alpha ^{\prime },$ non-zero dilaton and zero
axion}

Now we add the $\alpha ^{\prime }$ terms to the equations (2.2)-(2.3), but
neglect the axion. The resulting equations are reduced to three polynomial
constraints, 
\begin{eqnarray}
2\Omega ^2 &-&2\alpha ^{\prime }\Omega ^4=0, \\
2\Omega ^2 &-&m^2+\alpha ^{\prime }\left( 10\Omega ^4-6\Omega
^2m^2+m^4\right) =0, \\
2\Omega ^2 &-&2m^2-2\Lambda -f^2+\alpha ^{\prime }\left( 11\Omega ^4-6\Omega
^2m^2+m^4\right) =0.
\end{eqnarray}
We have three equations and five constants and so two (say, $\Lambda $ and $%
f $) can be chosen arbitrarily.

\mbox{$>$}
>From the equation (3.14) we immediately have the simple relation 
\begin{equation}
\alpha ^{\prime }=\frac 1{\Omega ^2},
\end{equation}
which gives the velocity of rotation of the G\"odel universe in terms of the
inverse string tension. This relation gives a simple connection between
micro and macrophysics in this spacetime, with a balance between string
tension and rotation. Using (3.17), subtracting (3.15) and (3.16), we find, 
\begin{equation}
m^2+2\Lambda +f^2=\Omega ^2.
\end{equation}
>From (3.15) and (3.17) we calculate the possible values of $\Omega $ in
terms of $m$ to be, \\
\begin{equation}
\Omega _{+}^2=\frac 13m^2,
\end{equation}
or 
\begin{equation}
\Omega _{+}^2=\frac 14m^2.
\end{equation}
The case (3.19) allows CTC's ($G(r)<0$ in (3.5)), just as in general
relativity. For the case (3.20), after using (3.16), we obtain the relations 
\begin{equation}
\Omega ^2=\frac 1{\alpha ^{\prime }}=\frac{m^2}4=\frac{-2\Lambda -f^2}3,
\end{equation}
and the condition (3.12) must still hold. This can also be related to the
number of dimensions $n=4$, using (2.6) to remove $\Lambda $ from the
relation (3.21), so 
\begin{equation}
\Omega ^2=\frac 1{\alpha ^{\prime }}=\frac{m^2}4=\frac 3{35}f^2,
\end{equation}
and the cosmological constant (2.6) must be negative, in agreement with \cite
{accio}.

\subsection{ First order in $\alpha ^{\prime }$, non-zero axion and dilaton}

In the case of the bosonic string ($\lambda _0=-1/4$), using (2.5)-(2.6),
the field equations (2.2)-(2.3) for the G\"odel metric reduce to the three
polynomials, 
\begin{eqnarray}
2\Omega ^2 &-&\frac 12E^2+\frac 12\alpha ^{\prime }\left[ -4\Omega
^4+4\Omega ^2E^2+\frac 54E^4\right] =0, \\
2\Omega ^2 &-&m^2+\frac 12E^2+\frac 12\alpha ^{\prime }\left[ 20\Omega
^4-12\Omega ^2m^2+2m^4-2E^2\left( m^2-2\Omega ^2\right) -\frac 54E^4\right]
=0, \\
2\Omega ^2 &-&2m^2-2\Lambda -f^2+\frac 12E^2+\frac 14\alpha ^{\prime }\left[
44\Omega ^4-24\Omega ^2m^2+4m^4-2E^2\left( m^2-\Omega ^2\right) -\frac 54%
E^4\right] =0.
\end{eqnarray}
The axion equation of motion (2.4) is also fulfilled in this general case.
We now have three equations and six constants, leaving three (say, $\Lambda $%
, $f$ and $E$) arbitrary.

\mbox{$>$}
From (3.23) and (3.24) we obtain 
\begin{equation}
\Omega ^2=\frac{m^2}4.
\end{equation}
This confirms that it is possible to obtain a G\"odel solution with no CTCs
which fulfils (3.5) in the general case. The value of $\alpha ^{\prime }$
can now be expressed in terms of the velocity of rotation of the universe, $%
\Omega ,$ and the strength of axion field, $E$, from (3.23), which gives, 
\begin{equation}
\alpha ^{\prime }=\frac{4\Omega ^2-E^2}{\left( 4\Omega ^2+E^2\right) \left(
\Omega ^2E-\frac 54E^2\right) }.
\end{equation}
The relation between other constants can be obtained from (3.25). Similar
calculations can also be performed for heterotic strings (with $\lambda _0=-%
\frac 18$ in (2.1)) which demonstrate the existence of G\"odel solutions
without CTCs within that theory also. The superstring case $(\lambda _0=0)$
does not generate a solution of this type to this order in $\alpha ^{\prime }
$ because the quadratic correction term vanishes in the action (2.1). The
persistence of the causal relation (3.26) between the free parameters in the
G\"odel metric suggests that it might continue to hold in solutions to
higher-orders but then we might expect the relation between $\alpha ^{\prime
}$ and $m$ to change, reflecting the contributions from the scales of the
higher-order terms in the action. This will be investigated elsewhere.

\section{Discussion}

We have found a class of G\"odel universes without closed timelike curves
within the framework of low-energy-effective string theories. First, to
zeroth order in the inverse string tension $\alpha ^{\prime }$, we
investigated the situation in both the Einstein frame (as studied already in
general relativity in \cite{tiomno}) and in the string frame. We found that
the axion cannot be introduced in the Einstein frame but plays a crucial
role in the string frame, where it cannot be neglected. Then we extended the
analysis to include the full $O(\alpha ^{\prime })$ action with both dilaton
and axion taken into account. By including terms of the first order in $%
\alpha ^{\prime }$ in the field equations, we found that G\"odel universes
without closed timelike curves are also possible in the general case. Our
solutions display a simple relation between the inverse string tension
parameter $\alpha ^{\prime }$ and the velocity of rotation of the universe
(eq. (3.17)) which provides a direct link between micro and macrophysics.
This is the first class of exact solutions that has been found for a string
theory where terms of first order in $\alpha ^{\prime }$ are admitted in the
field equations (a method for generating exact solutions in Bianchi type I
universes was given by Mueller in Ref.\cite{firstalfa}). All the G\"odel
universes we have found require negative cosmological constant which, in
general, can be related to the number of spacetime dimensions if
multidimensional cases are also investigated. Our results, obtained for
bosonic strings, are also valid for heterotic and superstrings, as is
evident from the form of the action, although the numerical values of the
constants will change with the value of the parameter $\lambda _0$. The
simplicity of the final forms of the solutions we have found for G\"odel
universes suggests that the symmetries of string theory have the power to
exclude unwanted peculiarities in the causal structure of spacetime that are
permitted by general relativity. They also suggest a path towards finding
further exact solutions at higher order in the string tension parameter.

\section{Acknowledgments}

MPD acknowledges the support of NATO and the Royal Society while at the
University of Sussex. JDB is supported by a PPARC Senior Fellowship.

\end{document}